\documentclass[12pt]{article}
\usepackage{amssym}

\def\C{\mbox{$\Bbb C$}}
\def\Qb{\bar{Q}}
\def\Ab{\bar{A}}
\def\cosech{\mathop{\rm cosech}\nolimits}
\def\sech{\mathop{\rm sech}\nolimits}
\def\case#1#2{{\textstyle{#1\over #2}}}
\def\Ap{\mbox{${\cal A}^+$}}
\def\Am{\mbox{${\cal A}^-$}}

\title{
\hfill{\normalsize ULB/229/CQ/01/3}\\
\vspace{1cm} 
COMPLEXIFIED PSUSY AND SSUSY INTERPRETATIONS OF SOME PT-SYMMETRIC
HAMILTONIANS POSSESSING TWO SERIES OF REAL ENERGY EIGENVALUES}
\author{B. BAGCHI\thanks{E-mail: bbagchi@cucc.ernet.in} \ and S.
MALLIK \\
{\small \sl Department of Applied Mathematics, University of Calcutta,} \\
{\small \sl 92 Acharya Prafulla Chandra Road, Calcutta 700 009, India}\\[10pt] 
C. QUESNE\thanks{Directeur de recherches FNRS; E-mail: cquesne@ulb.ac.be} \\
{\small \sl Physique Nucl\'eaire Th\'eorique et Physique Math\'ematique,}\\ {\small \sl
Universit\'e Libre de Bruxelles, Campus de la Plaine CP229,} \\ {\small \sl  Boulevard~du
Triomphe, B-1050 Brussels, Belgium}}
\date{ }
\begin{document}
\baselineskip=22pt plus 1pt minus 1pt
\maketitle

\begin{abstract} 
We analyze  a set of three PT-symmetric complex potentials, namely harmonic oscillator,
generalized P\"oschl-Teller and Scarf~II, all of which reveal a double series of energy
levels along with the corresponding superpotential. Inspired by the fact that two
superpotentials reside naturally in order-two parasupersymmetry (PSUSY) and
second-derivative supersymmetry (SSUSY) schemes, we complexify their frameworks to
successfully account for the three potentials.  
\end{abstract}

\vspace{1cm}
\noindent
Running head: Complexified PSUSY and SSUSY

\newpage
%
%
\section{Introduction}

In the literature, non-Hermitian Hamiltonians have received attention~\cite{feshbach} from
time to time because of their potential applications in scattering problems. Lately, a
subclass of such Hamiltonians, containing operators invariant under joint actions of parity
(P: $x \to -x$) and time reversal (T: ${\rm i} \to - {\rm i}$), has become a subject matter
of considerable research interest~\cite{bender98a}--\cite{dorey}. An important reason for
this is that PT invariance, in a number of cases, leads to energy eigenvalues that are real.
Moreover, PT-invariant models share with the usual Hermitian ones many of the features
that the latter admit of: namely, supersymmetrization~\cite{andrianov99, bagchi00a,
znojil00}, potential algebra~\cite{bagchi00c, levai01},
quasi-solvability~\cite{bender98b, znojil99a, bagchi00b, cannata}, etc.\par
%
%
Recently, Znojil~\cite{znojil99b}, by employing a simple complex shift of coordinate,
pointed out that the PT-symmetric harmonic oscillator potential possesses two series of
energy levels distinguishable by a quasi-parity parameter. Subsequently, we have also
found~\cite{bagchi00c} in an sl(2, \C) group theoretical context, that paired real energy
levels exist for a PT-symmetric generalized P\"oschl-Teller potential. The complexified
Scarf~II potential, which is also PT symmetric and emerges from the same sl(2, \C) algebra,
displays a double tower of real energy levels as well.\par
%
%
The purpose of this paper is to bring together these potentials within the framework of an
order-two parasupersymmetric (PSUSY) scheme and consequently interpret them in a
second-derivative supersymmetric (SSUSY) setting. In the Hermitian context, both the
procedures admit of two superpotentials. By complexifying them, we show that all the
three potentials mentioned above come under the purview of PSUSY and SSUSY. In this
way we establish that both PSUSY and SSUSY appear to be the most natural choice for
describing occurrences of a double series of energy levels.\par
%
%
\section{In pursuit of a complexified PSUSY}

\subsection{Underlying ideas of SUSY and PSUSY}

The basic principles of SUSY~\cite{cooper, bagchi00d} and PSUSY~\cite{rubakov,
beckers} in quantum mechanics~(QM) are well known. In SUSYQM, the governing
Hamiltonian is written in terms of a pair of supercharges $Q$ and $\Qb$, namely
\begin{equation}
  H_s = Q \Qb + \Qb Q. \label{eq:Hs}
\end{equation}
These supercharges are nilpotent and commute with $H_s$:
\begin{equation}
  Q^2 = \Qb^2 = 0, \qquad [H_s, Q] = [H_s, \Qb] = 0.
\end{equation}
The key role of $Q$ ($\Qb$) is that it operates on a bosonic state to transform it into a
fermionic one and vice versa.\par
%
%
In the minimal version of SUSY~\cite{witten}, $Q$ and $\Qb$ are generally assumed to
be represented by $Q = A \sigma_-$, $\Qb = \Ab \sigma_+$, where $A$ and $\Ab$ are
taken to be first-derivative differential operators. So one works with
\begin{eqnarray}
  Q & = & \left(\begin{array}{cc}
                0 & 0 \\
                A & 0
           \end{array}\right), \qquad
  \Qb = \left(\begin{array}{cc}
                0 & \Ab \\
                0 & 0
           \end{array}\right), \label{eq:SUSY-charges} \\
  A & = & \frac{d}{dx} + W(x), \qquad \Ab = - \frac{d}{dx} + W(x),
\end{eqnarray}
where $W(x)$ is the so-called superpotential. It is obvious from the above
representations of $Q$ and $\Qb$ that $H_s$ appears diagonal:
\begin{equation}
  H_s = \left(\begin{array}{cc}
                H_+ & 0 \\
                0 & H_-
           \end{array}\right). 
\end{equation}
\par
%
%
We can actually express $H_+$ and $H_-$ in factorized forms in terms of $A$ and
$\Ab$,
\begin{equation}
  H_+ = \Ab A = - \frac{d^2}{dx^2} + V_+(x) - E, \qquad H_- = A \Ab = -
  \frac{d^2}{dx^2} + V_-(x) - E, \label{eq:H-components} 
\end{equation}
at some arbitrary factorization energy $E$. In~(\ref{eq:H-components}), $V_{\pm}(x)$
are 
\begin{equation}
  V_{\pm} (x) = W^2(x) \mp \frac{dW(x)}{dx} + E. \label{eq:V-components} 
\end{equation}
\par
%
%
It may be noticed that the spectrum of $H_s$ is doubly degenerate except possibly for
the ground state. In the exact SUSY case to which we shall restrict ourselves here, the
ground state at vanishing energy is nondegenerate. In the present notational set-up, it
belongs to the $H_+$ component. Note that the double degeneracy of~$H_s$ is
also implied by the intertwining relationships, which read
\begin{equation}
  A H_+ = H_- A, \qquad H_+ \Ab = \Ab H_-. \label{eq:SUSY-intertwine} 
\end{equation}
Relations (\ref{eq:SUSY-intertwine}) are indeed consistent with the
definitions~(\ref{eq:H-components}).\par
%
%
PSUSY of order two ($p=2$), on the other hand, arises by imposing a symmetry between
the standard bosonic and parafermionic states. As introduced by Rubakov and
Spiridonov~\cite{rubakov}, the $p=2$ PSUSY Hamiltonian $H_{ps}$ is defined to obey
the relations
\begin{equation}
  Q^3 = 0, \qquad Q^2 \Qb + Q \Qb Q + \Qb Q^2 = 2Q H_{ps}, \qquad [H_{ps}, Q] = 0,
  \label{eq:PSUSY-alg}
\end{equation}
along with their Hermitian conjugates.\par
%
%
In parallel to (\ref{eq:SUSY-charges}), the parasupercharges $Q$ and $\Qb$ can be
assigned a matrix representation in a manner
\begin{equation}
  (Q)_{ij} = \left[\frac{d}{dx} + W_j(x)\right] \delta_{i, j+1}, \quad (\Qb)_{ij} =
  \left[- \frac{d}{dx} + W_i(x)\right] \delta_{i+1,j}, \quad i, j = 1, 2, 3.
\end{equation}
These read explicitly
\begin{equation}
  Q = \left(\begin{array}{ccc}
                0 & 0 & 0 \\
                A_1 & 0 & 0 \\
                0 & A_2 & 0
           \end{array}\right), \qquad
  \Qb = \left(\begin{array}{ccc}
                0 & \Ab_1 & 0 \\
                0 & 0 & \Ab_2 \\
                0 & 0 & 0
           \end{array}\right), 
\end{equation}
with
\begin{equation}
  A_i = \frac{d}{dx} + W_i(x), \qquad \Ab_i = - \frac{d}{dx} + W_i(x), \qquad i = 1, 2.
\end{equation}
\par
%
%
The PSUSY algebra (\ref{eq:PSUSY-alg}) then leads to a diagonal form for~$H_{ps}$,
\begin{equation}
  H_{ps} = \left(\begin{array}{ccc}
                H_1 & 0 & 0 \\
                0 & H_2 & 0 \\
                0 & 0 & H_3
           \end{array}\right), \label{eq:Hps}
\end{equation}
provided
\begin{equation}
  A_1 \Ab_1 = \Ab_2 A_2 - c, \label{eq:PSUSY-cond}
\end{equation}
where $c$ is a constant. Translated in terms of the superpotentials,
Eq.~(\ref{eq:PSUSY-cond}) expands to
\begin{equation}
  W_2^2 - W_1^2 - \frac{dW_1}{dx} - \frac{dW_2}{dx} = c. \label{eq:PSUSY-constraint}  
\end{equation}
We thus have for $H_1$, $H_2$, and $H_3$,
\begin{eqnarray}
  H_1 & = & \Ab_1 A_1 + c_1, \nonumber \\
  H_2 & = & A_1 \Ab_1 + c_1 = \Ab_2 A_2 + c_2, \nonumber \\
  H_3 & = & A_2 \Ab_2 + c_2, \label{eq:PSUSY-Hcomp}
\end{eqnarray}
where the constants $c_1$ and $c_2$ satisfy $c_1 + c_2 = 0$ and $c_1 - c_2 =
c$.\par
%
%
In summary, it is clear that whereas SUSY involves a single superpotential $W(x)$, PSUSY
is described by two superpotentials $W_1(x)$ and $W_2(x)$. We now turn to the case 
of the PT-symmetric harmonic oscillator potential for a PSUSY analysis.\par
%
%
\subsection{PT-symmetric oscillator potential}

The Hamiltonian~\cite{znojil99b}
\begin{equation}
  H^{(\alpha)} = - \frac{d^2}{dx^2} + (x - {\rm i} \delta)^2 + \frac{\alpha^2 -
  \frac{1}{4}}{(x - {\rm i} \delta)^2}, \qquad \alpha > 0, \label{eq:osc-H} 
\end{equation}
is easily seen to be PT symmetric: it can be obtained from the usual three-dimensional
radial harmonic oscillator Hamiltonian by effecting a complex shift of coordinate $x \to x -
{\rm i} \delta$, $\delta > 0$. The operator $H^{(\alpha)}$ is beset with a centrifugal-like
core of strength $G = \alpha^2 - \frac{1}{4}$; nonetheless, the model proves to be
exactly solvable on the entire real line for any $\alpha > 0$ like the linear harmonic
oscillator (corresponding to $\alpha = 1/2$). Contrary to the latter, however, it has an
unequal spectrum,
\begin{equation}
  E^{(\alpha)}_{qn} = 4n + 2 - 2q \alpha, \qquad n = 0, 1, 2, \ldots, \label{eq:osc-spec}
\end{equation}
if $\alpha$ is not integer, which we shall assume here. In~(\ref{eq:osc-spec}), $q = \pm
1$ denotes the quasi-even ($+$) or quasi-odd ($-$) parity for the corresponding state. The
accompanying eigenfunctions are expressible in terms of the standard orthogonal Laguerre
polynomials:
\begin{equation}
  \psi^{(\alpha)}_{qn}(x) \propto e^{-\frac{1}{2}(x - {\rm i} \delta)^2} 
  (x - {\rm i} \delta)^{-q\alpha + \frac{1}{2}} L_n^{(-q \alpha)}[(x - {\rm i} \delta)^2].
\end{equation}
\par
%
%
Before taking up the PSUSY study, it is interesting to discuss some of the SUSY aspects
of~$H^{(\alpha)}$. We see from~(\ref{eq:H-components}) and~(\ref{eq:V-components})
that there can be two independent\footnote{We can think of additional supersymmetries
resulting from the choices
\begin{eqnarray}
  W^{\prime\prime(\alpha)} (x) & = & x - {\rm i} \delta + \frac{\alpha + \frac{1}{2}}{x -
        {\rm i} \delta}, \qquad E'' = - 2 \alpha, \nonumber \\
  W^{\prime\prime\prime(\alpha)} (x) & = & x - {\rm i} \delta - \frac{\alpha -
        \frac{1}{2}}{x - {\rm i} \delta}, \qquad E''' = 2 \alpha, \nonumber 
\end{eqnarray}
where $E''$ and $E'''$ are the factorization energies. However, these supersymmetries
are not new in that they can be obtained from $W^{(\alpha)}(x)$ and $W^{\prime
(\alpha)}(x)$ by the replacement $\alpha \to \alpha + 1$ or $\alpha \to \alpha - 1$.}
forms of the complex superpotentials associated with $H^{(\alpha)}$. These are
\begin{eqnarray}
  W^{(\alpha)} (x) & = & x - {\rm i} \delta + \frac{\alpha - \frac{1}{2}}{x - {\rm i} \delta},
        \qquad E = 2 - 2 \alpha, \label{eq:osc-super1} \\
  W^{\prime(\alpha)} (x) & = & x - {\rm i} \delta - \frac{\alpha + \frac{1}{2}}{x -
        {\rm i} \delta}, \qquad E' = 2 + 2 \alpha. \label{eq:osc-super2} 
\end{eqnarray}
In (\ref{eq:osc-super1}) and (\ref{eq:osc-super2}), $E$ and $E'$ stand for the
corresponding factorization energies.\par
%
%
Let us consider $W^{(\alpha)}(x)$ first. Using~(\ref{eq:H-components})
and~(\ref{eq:V-components}), it follows readily that
\begin{equation}
  V^{(\alpha)}_+(x) = V^{(\alpha)}(x), \qquad V^{(\alpha)}_-(x) = V^{(\alpha-1)}(x) +
  2, 
\end{equation}
where $\alpha > 1$ and $V^{(\alpha)}(x)$ represents the potential in~(\ref{eq:osc-H}).
Thus the partner Hamiltonians $H^{(\alpha)}_{\pm}$ acquire the forms
\begin{equation}
  H^{(\alpha)}_+ = H^{(\alpha)} - 2 + 2 \alpha, \qquad H^{(\alpha)}_- = H^{(\alpha-1)} 
  + 2\alpha.
\end{equation}
\par
%
%
{}Further, using the definitions $A = \frac{d}{dx} + W^{(\alpha)}(x)$ and $\Ab = -
\frac{d}{dx} + W^{(\alpha)}(x)$, it is straightforward to verify that the operator~$A$
annihilates the ground state $\psi^{(\alpha)}_{+0}$:
\begin{equation}
  A \psi^{(\alpha)}_{+0}(x) \propto \left(\frac{d}{dx} + x - {\rm i} \delta + \frac{\alpha -
  \frac{1}{2}}{x - {\rm i} \delta}\right) e^{-\frac{1}{2}(x - {\rm i} \delta)^2} (x - {\rm i}
  \delta)^{- \alpha + \frac{1}{2}} = 0. 
\end{equation}
So the spectra of $H_+$ and $H_-$ read
\begin{eqnarray}
  \mbox{Spectrum of $H^{(\alpha)}_+$:} && E^{(\alpha)}_{+n} - 2 + 2\alpha = 4n,
         \nonumber \\
  && E^{(\alpha)}_{-n} - 2 + 2\alpha = 4n + 4\alpha, \label{eq:osc-spec1} \\  
  \mbox{Spectrum of $H^{(\alpha)}_-$:} && E^{(\alpha-1)}_{+n} + 2\alpha = 4n + 4,
         \nonumber \\
  && E^{(\alpha-1)}_{-n} + 2\alpha = 4n + 4\alpha. \label{eq:osc-spec2}  
\end{eqnarray}
\par
%
%
If however we consider $W^{\prime\alpha}(x)$ along with $E'$ given
by~(\ref{eq:osc-super2}), then $V^{\prime(\alpha)}_+(x)$ and
$V^{\prime(\alpha)}_-(x)$ become
\begin{equation}
  V^{\prime(\alpha)}_+(x) = V^{(\alpha)}(x), \qquad V^{\prime(\alpha)}_-(x)
  = V^{(\alpha+1)}(x) + 2.
\end{equation}
As such the corresponding component Hamiltonians $H^{\prime(\alpha)}_+$ and
$H^{\prime(\alpha)}_-$ turn out to be
\begin{equation}
  H^{\prime(\alpha)}_+ = H^{(\alpha)} - 2 - 2\alpha, \qquad H^{\prime(\alpha)}_- =
  H^{(\alpha+1)} - 2\alpha.
\end{equation}
\par
%
%
The role of $W^{\prime(\alpha)}(x)$ is, however, quite different from
$W^{(\alpha)}(x)$: it is the excited state $\psi^{(\alpha)}_{-0}$ that is
annihilated by $A'$ ($\equiv \frac{d}{dx} + W^{\prime(\alpha)}(x)$):
\begin{equation}
  A' \psi^{(\alpha)}_{-0}(x) \propto \left(\frac{d}{dx} + x - {\rm i} \delta - \frac{\alpha +
  \frac{1}{2}}{x - {\rm i} \delta}\right) e^{-\frac{1}{2}(x - {\rm i} \delta)^2} (x - {\rm i}
  \delta)^{\alpha + \frac{1}{2}} = 0. 
\end{equation}
As a result, the spectra of $H^{\prime(\alpha)}_+$ and $H^{\prime(\alpha)}_-$ look
much different from those in~(\ref{eq:osc-spec1}) and~(\ref{eq:osc-spec2}):
\begin{eqnarray}
  \mbox{Spectrum of $H^{\prime(\alpha)}_+$:} && E^{(\alpha)}_{+n} - 2 - 2\alpha = 4n -
         4\alpha, \nonumber \\
  && E^{(\alpha)}_{-n} - 2 - 2\alpha = 4n, \\  
  \mbox{Spectrum of $H^{\prime(\alpha)}_-$:} && E^{(\alpha+1)}_{+n} - 2\alpha = 4n -
         4\alpha, \nonumber \\
  && E^{(\alpha+1)}_{-n} - 2\alpha = 4n + 4.
\end{eqnarray}
\par
%
%
With this background, we now proceed to discuss the PSUSY of the PT-symmetric
oscillator Hamiltonian~(\ref{eq:osc-H}). Introducing a pair of complex superpotentials
$W_1(x) = W^{(\alpha)}(x)$ and $W_2(x) = W^{\prime(\alpha-1)}(x)$ and taking $c_1
= - c_2 = - 2\alpha$, we at once obtain from~(\ref{eq:PSUSY-Hcomp}) the results:
\begin{equation}
  H_1 = H^{(\alpha)}_+ - 2\alpha = H^{(\alpha)} - 2, \qquad H_2 = H^{(\alpha)}_- -
  2\alpha = H^{(\alpha-1)}, \qquad H_3 = H^{(\alpha)} + 2. \label{eq:osc-para-H} 
\end{equation}
\par
%
%
We are therefore led to the following PSUSY spectrum pattern:
\begin{eqnarray}
  \mbox{Spectrum of $H_1$:} && E^{(\alpha)}_{+n} - 2 = 4n - 2\alpha,
         \nonumber \\
  && E^{(\alpha)}_{-n} - 2 = 4n + 2\alpha, \label{eq:osc-paraspec1} \\  
  \mbox{Spectrum of $H_2$:} && E^{(\alpha-1)}_{+n} = 4n - 2\alpha + 4,
         \nonumber \\
  && E^{(\alpha-1)}_{-n} = 4n + 2\alpha, \\  
  \mbox{Spectrum of $H_3$:} && E^{(\alpha)}_{+n} + 2 = 4n - 2\alpha + 4,
         \nonumber \\
  && E^{(\alpha)}_{-n} + 2 = 4n + 2\alpha + 4. \label{eq:osc-paraspec3}  
\end{eqnarray}
As a consequence of Eqs.~(\ref{eq:osc-paraspec1})--(\ref{eq:osc-paraspec3}), the
spectrum of $H_{ps}$ shows the features summarized below: if $N-1 < \alpha < N$,
where $N \in \{2, 3, \ldots\}$, then
\begin{eqnarray}
  E_0 & = & - 2\alpha, \quad E_1 = - 2\alpha + 4, \quad \ldots, E_{N-1} = - 2\alpha + 4N
       - 4, \quad E_N = 2\alpha, \nonumber \\
  E_{N+1} & = & - 2\alpha + 4N, \quad E_{N+2} = 2\alpha + 4, \quad \ldots, E_{N+2m} =
       2\alpha + 4m, \nonumber \\
  E_{N+2m+1} & = & - 2\alpha + 4N + 4m, \quad \ldots, 
\end{eqnarray}
with degeneracies
\begin{eqnarray}
  d_0 & = & 1, \quad d_1 = 3, \quad \ldots, d_{N-1} = 3, \quad d_N = 2, \quad d_{N+1}
       = 3, \quad d_{N+2} = 3, \quad \ldots, \nonumber \\
  d_{N+2m} & = & 3, \quad d_{N+2m+1} = 3, \quad \dots. \label{eq:osc-para-d}
\end{eqnarray}
\par
%
%
In the Hermitian case, the spectrum of $H_{ps}$ is known to be always three-fold
degenerate at least starting from the second and higher excited states.
From~(\ref{eq:osc-para-d}), we see that this is not true here. Note that $H_3$ is
essentially a shifted $H_1$ and that the ground state is nondegenerate.\par
%
%
We now remark on the other possibility when we can identify $W'_1(x) =
W^{\prime(\alpha)}(x)$ and $W'_2(x) = W^{(\alpha+1)}(x)$ with $c'_1 = - c'_2 =
2\alpha$. We obtain as a result
\begin{equation}
  H'_1 = H^{\prime(\alpha)}_+ + 2\alpha = H^{(\alpha)} - 2, \qquad H'_2 =
  H^{\prime(\alpha)}_- + 2\alpha = H^{(\alpha+1)}, \qquad H'_3 = H^{(\alpha)} + 2. 
  \label{eq:osc-para-Hbis}
\end{equation}
The respective spectra of $H'_1$, $H'_2$, and $H'_3$ are then
\begin{eqnarray}
  \mbox{Spectrum of $H'_1$:} && E^{(\alpha)}_{+n} - 2 = 4n - 2\alpha,
         \nonumber \\
  && E^{(\alpha)}_{-n} - 2 = 4n + 2\alpha, \\  
  \mbox{Spectrum of $H'_2$:} && E^{(\alpha+1)}_{+n} = 4n - 2\alpha,
         \nonumber \\
  && E^{(\alpha+1)}_{-n} = 4n + 2\alpha + 4, \\  
  \mbox{Spectrum of $H'_3$:} && E^{(\alpha)}_{+n} + 2 = 4n - 2\alpha + 4,
         \nonumber \\
  && E^{(\alpha)}_{-n} + 2 = 4n + 2\alpha + 4. 
\end{eqnarray}
\par
%
%
These yield the following spectrum of $H_{ps}$: if $N-1 < \alpha < N$, where $N \in \{1,
2,
\ldots\}$, then
\begin{eqnarray}
  E_0 & = & - 2\alpha, \quad E_1 = - 2\alpha + 4, \quad \ldots, E_{N-1} = - 2\alpha + 4N
       - 4, \quad E_N = 2\alpha, \nonumber \\
  E_{N+1} & = & - 2\alpha + 4N, \quad E_{N+2} = 2\alpha + 4, \quad \ldots, E_{N+2m} =
       2\alpha + 4m, \nonumber \\
  E_{N+2m+1} & = & - 2\alpha + 4N + 4m, \quad \ldots, 
\end{eqnarray}
with degeneracies
\begin{eqnarray}
  d_0 & = & 2, \quad d_1 = 3, \quad \ldots, d_{N-1} = 3, \quad d_N = 1, \quad d_{N+1}
       = 3, \quad d_{N+2} = 3, \quad \ldots, \nonumber \\
  d_{N+2m} & = & 3, \quad d_{N+2m+1} = 3, \quad \dots.
\end{eqnarray}
\par
%
%
In contrast to the previous case, here the ground state is doubly degenerate. However,
similar to what we obtained before, the nature of degeneracies is not of the usual
Hermitian type. Note that $H_3$ is again a shifted $H_1$. We
therefore conclude that to get a shifted PT-symmetric oscillator, one has to resort to a
complexified PSUSY of order two, contrary to what happens for the standard harmonic
oscillator case where such a result is obtained in SUSYQM.\par
%
%
In the limiting cases where $\alpha$ becomes some integer $N$, the spectrum of
$H^{(\alpha)}$ becomes equidistant and the need for quasi-parity disappears due to the
phenomenon of unavoided level crossings without degeneracy~\cite{znojil99b}. We then
recover a degeneracy pattern of the usual Hermitian type for the spectrum of $H_{ps}$,
namely
\begin{equation}
  d_0 = 1, \quad d_1 = 3, \quad \ldots, d_{N-1} = 3, \quad d_N = 3, \quad d_{N+1} = 3,
  \quad d_{N+2} = 3, \quad \ldots, 
\end{equation}
or
\begin{equation}
  d_0 = 2, \quad d_1 = 3, \quad \ldots, d_{N-1} = 3, \quad d_N = 3, \quad d_{N+1} = 3,
  \quad d_{N+2} = 3, \quad \ldots,
\end{equation}
for the choice~(\ref{eq:osc-para-H}) or~(\ref{eq:osc-para-Hbis}), respectively. In both
cases, the spectrum of $H_{ps}$ is the same:
\begin{eqnarray}
  E_0 & = & - 2N, \quad E_1 = - 2N + 4, \quad \ldots, E_{N-1} = 2N - 4, \quad E_N =
       2N, \nonumber \\ 
  E_{N+1} & = & 2N + 4, \quad E_{N+2}  = 2N + 8, \quad \ldots,
\end{eqnarray}
but for the former choice, $N$ is restricted to the set $\{2, 3, 4, \ldots\}$, while for the
latter it may take any value in $\{1, 2, 3, \ldots\}$. \par
%
%
\subsection{PT-symmetric generalized P\"oschl-Teller potential}

The Hamiltonian for the PT-symmetric generalized P\"oschl-Teller system is given
by~\cite{bagchi00c}
\begin{equation}
  H^{(A,B)} = - \frac{d^2}{dx^2} + [B^2 + A (A+1)] \cosech^2 \tau - B (2A+1) \cosech
  \tau \coth \tau, \qquad \tau = x - {\rm i} \gamma, \label{eq:PT-H} 
\end{equation}
where $- \frac{\pi}{4} \le \gamma < 0$ or $0 < \gamma < \frac{\pi}{4}$, $B > A +
\frac{1}{2} > 0$, and $A + \frac{1}{2}$ and $B$ do not differ by an integer. Note that
$H^{(A,B)}$ is invariant under the replacements $\left(A + \frac{1}{2}, B\right) \to
\left(B, A + \frac{1}{2}\right)$.\par
%
%
We have recently shown~\cite{bagchi00c}, using sl(2,\C) as a tool, that the
PT-symmetric Hamiltonian $H^{(A,B)}$ possesses two series of real energy eigenvalues
according to
\begin{eqnarray}
  E^{(A,B)}_{+n} & = & - \left(B - \case{1}{2} - n\right)^2, \qquad n = 0, 1, \ldots,
        n_{+ max}, \nonumber \\
  && B - \case{3}{2} \le n_{+ max} < B - \case{1}{2}, \label{eq:PT-E1} \\
  E^{(A,B)}_{-n} & = & - \left(A - n\right)^2, \qquad n = 0, 1, \ldots, n_{- max},
        \nonumber \\
  && A - 1 \le n_{- max} < A, \label{eq:PT-E2} 
\end{eqnarray}
where $B > \frac{1}{2}$ and $A > 0$. Note that while the real counterpart
of~(\ref{eq:PT-H}), obtained by setting $\gamma = 0$, is singular and so calls for its
restriction to the half-line $(0, + \infty)$, the complexified potential as given above gets
regularized on performing the shift $x \to x - {\rm i} \gamma$ and so may be considered
on the entire real line. Note also that the coupling constants appearing in $H^{(A,B)}$
are all real.\par
%
%
Corresponding to the two series of energy levels~(\ref{eq:PT-E1}) and~(\ref{eq:PT-E2}),
the eigenfunctions read
\begin{eqnarray}
  \psi^{(A,B)}_{+n} & \propto & (y-1)^{(A-B+1)/2} (y+1)^{-(A+B)/2}
        P_n^{(A-B+\frac{1}{2}, -A-B-\frac{1}{2})}(y), \\
  \psi^{(A,B)}_{-n} & \propto & (y-1)^{(B-A)/2} (y+1)^{-(B+A)/2}
        P_n^{(B-A-\frac{1}{2}, -B-A-\frac{1}{2})}(y),
\end{eqnarray}
where $y = \cosh \tau$ and $P^{(\alpha, \beta)}_n(y)$ is a Jacobi polynomial.\par
%
%
Carrying out a standard SUSYQM analysis, we get for $A = \frac{d}{dx} + W^{(A,B)}(x)$
and $\Ab = - \frac{d}{dx} + W^{(A,B)}(x)$ the partner Hamiltonians
\begin{eqnarray}
  H^{(A,B)}_+ & = & \Ab A = - \frac{d^2}{dx^2} + V^{(A,B)}_+(x) - E, \nonumber \\
  H^{(A,B)}_- & = & A \Ab = - \frac{d^2}{dx^2} + V^{(A,B)}_-(x) - E, 
\end{eqnarray}
where $V^{(A,B)}_{\pm}(x)$ are related to $W^{(A,B)}(x)$ as defined
in~(\ref{eq:V-components}). The superpotential $W^{(A,B)}$ is given by
\begin{equation}
  W^{(A,B)}(x) = \left(B - \case{1}{2}\right) \coth \tau - \left(A + \case{1}{2}\right)
  \cosech \tau, \qquad E = - \left(B - \case{1}{2}\right)^2, 
\end{equation}
$E$ being the factorization energy.\par
%
%
It is simple to work out
\begin{equation}
  V^{(A,B)}_+(x) = V^{(A,B)}(x), \qquad V^{(A,B)}_-(x) = V^{(A,B-1)}(x),
  \label{eq:PT-V-comp}
\end{equation}
where $V^{(A,B)}$ is the potential in~(\ref{eq:PT-H}). Relations~(\ref{eq:PT-V-comp})
imply as a consequence
\begin{equation}
  H^{(A,B)}_+ = H^{(A,B)} + \left(B - \case{1}{2}\right)^2, \qquad   H^{(A,B)}_- =
  H^{(A,B-1)} + \left(B - \case{1}{2}\right)^2. \label{eq:PT-H-comp}
\end{equation}
\par
%
%
The nondegenerate ground state $\psi^{(A,B)}_{+0}$ is easily seen to be annihilated by
the operator~$A$:
\begin{eqnarray}
  A \psi^{(A,B)}_{+0} & \propto & \left[\frac{d}{dx} + \left(B - \frac{1}{2}\right) \coth
         \tau - \left(A + \frac{1}{2}\right) \cosech \tau\right] (y-1)^{(A-B+1)/2}
         \nonumber \\ 
  && \mbox{} \times (y+1)^{-(A+B)/2} \nonumber \\ 
  & \propto & (\sinh \tau)^{-1} \left[\left(y^2 - 1\right) \frac{d}{dy} + \left(B -
         \frac{1}{2}\right) y - \left(A + \frac{1}{2}\right)\right] (y-1)^{(A-B+1)/2}
         \nonumber \\ 
  && \mbox{} \times (y+1)^{-(A+B)/2} \nonumber \\  
  & = & 0,
\end{eqnarray}
resulting in the following spectra of $H^{(A,B)}_{\pm}$:
\begin{eqnarray}
  \mbox{Spectrum of $H^{(A,B)}_+$:} && E^{(A,B)}_{+n} + \left(B -
        \case{1}{2}\right)^2 = n (2B - n -1), \nonumber \\
  && E^{(A,B)}_{-n} + \left(B - \case{1}{2}\right)^2 = \left(B - A + n - \case{1}{2}
        \right) \nonumber \\
  && \mbox{} \times \left(B + A - n - \case{1}{2}
        \right), \label{eq:PT-spec1} \\  
  \mbox{Spectrum of $H^{(A,B)}_-$:} && E^{(A,B-1)}_{+n} + \left(B -
        \case{1}{2}\right)^2 = (n + 1) (2B - n -2), \nonumber \\
  && E^{(A,B-1)}_{-n} + \left(B - \case{1}{2}\right)^2 = \left(B - A + n - \case{1}{2}
        \right) \nonumber \\
  && \mbox{} \times \left(B + A - n - \case{1}{2}
        \right). \label{eq:PT-spec2}
\end{eqnarray}
Clearly from (\ref{eq:PT-spec1}) and (\ref{eq:PT-spec2}) we get the usual picture of
unbroken SUSY.\par
%
%
Since the potential $V^{(A,B)}(x)$ is invariant under $A + \frac{1}{2} \leftrightarrow B$,
we may as well have a second choice of the superpotential given by
\begin{equation}
  W^{\prime(A,B)}(x) = A \coth \tau - B \cosech \tau, \qquad E' = - A^2,
\end{equation}
where $E'$ is the factorization energy. In this case, the wave function
$\psi^{(A,B)}_{-0}$ is annihilated by the operator~$A'$, showing that an excited state
at vanishing energy is suppressed:
\begin{eqnarray}
  \mbox{Spectrum of $H^{\prime(A,B)}_+$:} && E^{(A,B)}_{+n} + A^2 = 
        \left(A -B + n + \case{1}{2}\right) \left(A + B - n - \case{1}{2}
        \right), \nonumber \\
  && E^{(A,B)}_{-n} + A^2 = n (2A - n), \\
  \mbox{Spectrum of $H^{\prime(A,B)}_-$:} && E^{(A-1,B)}_{+n} + A^2 = 
        \left(A -B + n + \case{1}{2}\right) \left(A + B - n - \case{1}{2}
        \right), \nonumber \\
  && E^{(A-1,B)}_{-n} + A^2 = (n + 1) (2A - n - 1).
\end{eqnarray}
\par
%
%
Moving on to PSUSY we consider, as a first choice, the superpotentials $W_1(x)$ and
$W_2(x)$ defined by
\begin{eqnarray}
  W_1(x)  & = & W^{(A,B)}(x), \qquad W_2(x) = W^{\prime(A,B-1)}(x), \nonumber \\
  c_1 & = & - c_2
  = \case{1}{2} \left[A^2 - \left(B - \case{1}{2}\right)^2\right].
\end{eqnarray}
We then get for the component Hamiltonians of $H_{ps}$,
\begin{eqnarray}
  H_1 & = & H^{(A,B)} + {\cal E}, \qquad H_2 = H^{(A,B-1)} + {\cal E}, \qquad H_3 =
         H^{(A-1,B-1)} + {\cal E}, \nonumber \\
  {\cal E} & \equiv & \case{1}{2} \left[A^2 + \left(B - \case{1}{2}\right)^2\right],  
         \label{eq:PT-paracomp}
\end{eqnarray}
where (\ref{eq:PT-H-comp}) has been used. As a result, the following spectra of $H_1$,
$H_2$, and $H_3$ emerge:
\begin{eqnarray}
  \mbox{Spectrum of $H_1$:} && E^{(A,B)}_{+n} + {\cal E} = - \left(B - \case{1}{2} - n
        \right)^2 + {\cal E}, \nonumber \\
  && E^{(A,B)}_{-n} + {\cal E} = - \left(A - n\right)^2 + {\cal E}, \\
  \mbox{Spectrum of $H_2$:} && E^{(A,B-1)}_{+n} + {\cal E} = - \left(B - \case{3}{2} -
        n\right)^2 + {\cal E}, \nonumber \\
  && E^{(A,B-1)}_{-n} + {\cal E} = - \left(A - n\right)^2 + {\cal E},  \\
  \mbox{Spectrum of $H_3$:} && E^{(A-1,B-1)}_{+n} + {\cal E} = - \left(B - \case{3}{2}
        - n\right)^2 + {\cal E}, \nonumber \\
  && E^{(A-1,B-1)}_{-n} + {\cal E} = - \left(A - 1 - n\right)^2 + {\cal E}.
\end{eqnarray}
\par
%
%
We therefore see that in going from $H_1$ to $H_2$, one suppresses the ground state
of $H_1$ at an energy $\case{1}{2} \left[A^2 - \left(B - \case{1}{2}\right)^2\right] <
0$. Then in going from $H_2$ to $H_3$, one suppresses a state of $H_2$ at an energy 
$\case{1}{2} \left[\left(B - \case{1}{2}\right)^2 - A^2\right] > 0$. The latter is either
an excited state or the ground state according to whether $B > A + \case{3}{2}$ or $B <
A + \case{3}{2}$.\par
%
%
{}For completeness, let us write down the spectrum of $H_{ps}$. If $B - N < A +
\case{1}{2} < B - N + 1$, where $N \in \{1, 2, 3, \ldots\}$, it reads
\begin{eqnarray}
  E_0 & = & \case{1}{2} \left[A^2 - \left(B - \case{1}{2}\right)^2\right], \qquad d_0 =
       1, \nonumber \\
  E_1 & = & E_0 + 2B - 2, \qquad d_1 = 3, \nonumber \\
  & \vdots & \nonumber \\
  E_{N-1} & = & E_0 + (N - 1)(2B - N), \qquad d_{N-1} = 3, \nonumber \\
  E_N & = & E_0 - A^2 + \left(B - \case{1}{2}\right)^2, \qquad d_N = 2, \nonumber \\
  & \vdots & \nonumber \\
  E_{N+2p+1} & = & E_0 + (N + p)(2B - N - p - 1), \qquad d_{N+2p+1} = 3, \nonumber
        \\
  && p = 0, 1, \ldots, p_{+max}, \nonumber \\
  E_{N+2p} & = & E_0 - (A - p)^2 + \left(B - \case{1}{2}\right)^2, \qquad d_{N+2p} = 3,
        \nonumber \\
  && p = 1, 2, \ldots, p_{-max}. \label{eq:PT-paraspec}
\end{eqnarray} 
In (\ref{eq:PT-paraspec}), $d_i$ ($i = 0$, 1, \ldots, $N-1$, $N$, \ldots, $N+2p+1$,
$N+2p$) is the degeneracy, $p_{+max} = n_{+max} - N$, and $p_{-max} =
n_{-max}$.\par
%
%
We next consider the second choice of the superpotentials, namely
\begin{eqnarray}
  W'_1(x)  & = & W^{\prime(A,B)}(x), \qquad W'_2(x) = W^{(A-1,B)}(x),
         \nonumber \\
  c'_1 & = & - c'_2 = \case{1}{2} \left[\left(B - \case{1}{2}\right)^2 - A^2\right].
\end{eqnarray}
We obtain after a little algebra
\begin{equation}
  H'_1 = H^{(A,B)} + {\cal E}, \qquad H'_2 = H^{(A-1,B)} + {\cal E}, \qquad H'_3 =
         H^{(A-1,B-1)} + {\cal E},
\end{equation}
where $\cal E$ is the same as in~(\ref{eq:PT-paracomp}). The spectra of $H'_1$,
$H'_2$, and $H'_3$ read
\begin{eqnarray}
  \mbox{Spectrum of $H'_1$:} && E^{(A,B)}_{+n} + {\cal E} = - \left(B - \case{1}{2} - n
        \right)^2 + {\cal E}, \nonumber \\
  && E^{(A,B)}_{-n} + {\cal E} = - \left(A - n\right)^2 + {\cal E}, \\
  \mbox{Spectrum of $H'_2$:} && E^{(A-1,B)}_{+n} + {\cal E} = - \left(B - \case{1}{2} -
        n\right)^2 + {\cal E}, \nonumber \\
  && E^{(A-1,B)}_{-n} + {\cal E} = - \left(A - 1 - n\right)^2 + {\cal E}, \\
  \mbox{Spectrum of $H'_3$:} && E^{(A-1,B-1)}_{+n} + {\cal E} = - \left(B - \case{3}{2}
        - n\right)^2 + {\cal E}, \nonumber \\
  && E^{(A-1,B-1)}_{-n} + {\cal E} = - \left(A - 1 - n\right)^2 + {\cal E}.
\end{eqnarray}
\par
%
%
We thus see that in going from $H'_1$ to $H'_2$, one suppresses an excited state
of $H'_1$ at an energy $\case{1}{2} \left[\left(B - \case{1}{2}\right)^2 - A^2\right] >
0$. Then in going from $H'_2$ to $H'_3$, one suppresses the ground state of $H'_2$ at
an energy $\case{1}{2} \left[A^2 - \left(B - \case{1}{2}\right)^2\right] < 0$. Further if
$B - N < A + \case{1}{2} < B - N + 1$, where $N \in \{1, 2, 3, \ldots\}$, then the
spectrum of $H_{ps}$ is the same as in the previous case, but the degeneracies are $d_0
= 2$,
$d_1 = 3$, \ldots, $d_{N-1} = 3$, $d_N = 1$, \ldots, $d_{N+2p+1} = 3$, $d_{N+2p} =
3$.\par
%
%
In the limiting cases where $A + \frac{1}{2}$ and $B$ differ by some integer,
$H^{(A,B)}$ has a single series of energy levels due to the phenomenon of unavoided
level crossings without degeneracy~\cite{bagchi00c}. The PSUSY scheme then becomes
similar to the usual one for Hermitian Hamiltonians.\par
%
%
\subsection{PT-symmetric Scarf II potential}

The Hamiltonian for the PT-symmetric Scarf II potential is given by~\cite{bagchi00c}
\begin{equation}
  H^{(A,B)} = - \frac{d^2}{dx^2} - \left[B^2 + A (A+1)\right] \sech^2 x + {\rm i} B
  (2A+1) \sech x \tanh x, \label{eq:Scarf-H}
\end{equation}
where $A > B - \case{1}{2} > 0$ and $A - B + \case{1}{2}$ is not an integer. The
form~(\ref{eq:Scarf-H}) is PT symmetric; like the PT-symmetric generalized
P\"oschl-Teller Hamiltonian~(\ref{eq:PT-H}), it also exhibits invariance under exchange of
the parameters $A + \case{1}{2}$ and $B$. Full and detailed analyses of the various
properties of~(\ref{eq:Scarf-H}) have already been given by us
elsewhere~\cite{bagchi00c} in connection with sl(2,\C) potential algebra. We have found
that PT-symmetric Scarf II potential depicts two series of energy levels. These are
\begin{eqnarray}
  E^{(A,B)}_{+n} & = & - \left(A - n\right)^2, \qquad n = 0, 1, \ldots,
        n_{+ max}, \nonumber \\
  && A - 1 \le n_{+ max} < A,  \\
  E^{(A,B)}_{-n} & = & - \left(B - \case{1}{2} - n\right)^2, \qquad n = 0, 1, \ldots, 
        n_{-max}, \nonumber \\
  && B - \case{3}{2} \le n_{- max} < B - \case{1}{2}.
\end{eqnarray}
The accompanying eigenfunctions read
\begin{eqnarray}
  \psi^{(A,B)}_{+n} & \propto & (\sech x)^A \exp[- {\rm i} B \arctan(\sinh x)]
        P_n^{(-A+B-\frac{1}{2}, -A-B-\frac{1}{2})}({\rm i} \sinh x), \\
  \psi^{(A,B)}_{-n} & \propto & (\sech x)^{B - \frac{1}{2}} \exp\left[- {\rm i} \left(A +
        \case{1}{2}\right)\arctan(\sinh x)\right] \nonumber \\
  && \mbox{} \times P_n^{(A-B+\frac{1}{2}, -A-B-\frac{1}{2})}({\rm i} \sinh x),
\end{eqnarray}
in terms of Jacobi polynomials.\par
%
%
We are now going to show that the model~(\ref{eq:Scarf-H}) possesses PSUSY. This can
be easily established, as we did for the PT-symmetric oscillator and generalized
P\"oschl-Teller potentials, by demonstrating first that two superpotentials exist for it in
the context of SUSY. PSUSY can then be contructed taking their help.\par
%
%
Indeed one can verify that two possible candidates of the superpotential are
\begin{eqnarray}
  W^{(A,B)}(x) & = & A \tanh x + {\rm i} B \sech x, \quad E = - A^2, 
        \label{eq:Scarf-super1}\\
  W^{\prime(A,B)}(x) & = & \left(B - \case{1}{2}\right) \tanh x + {\rm i} \left(A +
        \case{1}{2}\right) \sech x, \quad E' = - \left(B - \case{1}{2}\right)^2,
        \label{eq:Scarf-super2}
\end{eqnarray}
where $E$ and $E'$ are the factorization energies. Note that (\ref{eq:Scarf-super2}) is
obtainable from~(\ref{eq:Scarf-super1}) under the replacements $A + \case{1}{2}
\leftrightarrow B$.\par
%
%
While corresponding to~(\ref{eq:Scarf-super1}) we derive
\begin{equation}
  V^{(A,B)}_+(x) = V^{(A,B)}(x), \qquad V^{(A,B)}_-(x) = V^{(A-1,B)}(x),
  \label{eq:Scarf-V-comp1}
\end{equation}
where $V^{(A,B)}(x)$ is the potential of (\ref{eq:Scarf-H}), Eq.~(\ref{eq:Scarf-super2})
yields the pair
\begin{equation}
  V^{\prime(A,B)}_+(x) = V^{(A,B)}(x), \qquad V^{\prime(A,B)}_-(x) = V^{(A,B-1)}(x).
  \label{eq:Scarf-V-comp2}
\end{equation}
\par
%
%
The associated partner Hamiltonians for~(\ref{eq:Scarf-V-comp1})
and~(\ref{eq:Scarf-V-comp2}) are
\begin{eqnarray}
  H^{(A,B)}_+ & = & H^{(A,B)} + A^2, \qquad   H^{(A,B)}_- = H^{(A-1,B)} + A^2,
        \label{eq:Scarf-H-comp} \\
  H^{\prime(A,B)}_+ & = & H^{(A,B)} + \left(B - \case{1}{2}\right)^2, \qquad  
        H^{\prime(A,B)}_- = H^{(A,B-1)} + \left(B - \case{1}{2}\right)^2.
        \label{eq:Scarf-H-compbis}
\end{eqnarray}
Further, defining operators $A = \frac{d}{dx} + W^{(A,B)}(x)$ and $A' = \frac{d}{dx} +
W^{\prime(A,B)}(x)$, it is a simple exercise to check that the states
$\psi^{(A,B)}_{+0}$ and $\psi^{(A,B)}_{-0}$ are annihilated by $A$ and $A'$,
respectively. The spectra of $H^{(A,B)}_{\pm}$ and $H^{\prime(A,B)}_{\pm}$ turn
out to be
\begin{eqnarray}
  \mbox{Spectrum of $H^{(A,B)}_+$:} && E^{(A,B)}_{+n} + A^2 = n (2A - n),
        \nonumber \\
  && E^{(A,B)}_{-n} + A^2 = \left(A - B + n + \case{1}{2}
        \right) \nonumber \\
  && \mbox{} \times \left(A + B - n - \case{1}{2}\right), \label{eq:Scarf-spec1} \\  
  \mbox{Spectrum of $H^{(A,B)}_-$:} && E^{(A-1,B)}_{+n} + A^2 = (n + 1) (2A - n -1),
        \nonumber \\
  && E^{(A-1,B)}_{-n} + A^2 = \left(A - B + n + \case{1}{2}\right) \nonumber \\
  && \mbox{} \times \left(A + B - n - \case{1}{2}\right), \label{eq:Scarf-spec2}\\
  \mbox{Spectrum of $H^{\prime(A,B)}_+$:} && E^{(A,B)}_{+n} + \left(B -
        \case{1}{2}\right)^2 = \left(B - A + n - \case{1}{2} \right) \nonumber \\
  && \mbox{} \times \left(B + A - n - \case{1}{2}\right), \nonumber\\ 
  && E^{(A,B)}_{-n} + \left(B - \case{1}{2}\right)^2 = n (2B - n - 1),
        \label{eq:Scarf-spec1bis}\\  
  \mbox{Spectrum of $H^{\prime(A,B)}_-$:} && E^{(A,B-1)}_{+n} + \left(B -
        \case{1}{2}\right)^2 = \left(B - A + n - \case{1}{2} \right) \nonumber \\
  && \mbox{} \times \left(B + A - n - \case{1}{2}\right), \nonumber \\
  && E^{(A,B-1)}_{-n} + \left(B - \case{1}{2}\right)^2 = (n + 1)(2B - n - 2).
        \label{eq:Scarf-spec2bis}
\end{eqnarray}
While (\ref{eq:Scarf-spec1}) and (\ref{eq:Scarf-spec2}) show the conventional
unbroken SUSY picture, (\ref{eq:Scarf-spec1bis}) and (\ref{eq:Scarf-spec2bis}) point to
an unusual scenario: an excited state at vanishing energy is suppressed.\par
%
%
Equipped with the above SUSY machinery, we define the following pair of superpotentials
for $p=2$ PSUSY:
\begin{eqnarray}
  W_1(x)  & = & W^{(A,B)}(x), \qquad W_2(x) = W^{\prime(A-1,B)}(x), \nonumber \\
  c_1 & = & - c_2 = \case{1}{2} \left[\left(B - \case{1}{2}\right)^2 - A^2\right].
        \label{eq:Scarf-para-super}
\end{eqnarray}
Then it follows from~(\ref{eq:PSUSY-Hcomp}), (\ref{eq:Scarf-H-comp}),
and~(\ref{eq:Scarf-H-compbis}) that
\begin{eqnarray}
  H_1 & = & H^{(A,B)} + {\cal E}, \qquad H_2 = H^{(A-1,B)} + {\cal E}, \qquad H_3 =
         H^{(A-1,B-1)} + {\cal E}, \nonumber \\
  {\cal E} & \equiv & \case{1}{2} \left[A^2 + \left(B - \case{1}{2}\right)^2\right].  
\end{eqnarray}
The spectra of $H_1$, $H_2$, and $H_3$ are
\begin{eqnarray}
  \mbox{Spectrum of $H_1$:} && E^{(A,B)}_{+n} + {\cal E} = - \left(A - n\right)^2 +
        {\cal E}, \nonumber \\
  && E^{(A,B)}_{-n} + {\cal E} = - \left(B - \case{1}{2} - n
        \right)^2 + {\cal E}, \label{eq:Scarf-para-comp1}\\
  \mbox{Spectrum of $H_2$:} && E^{(A-1,B)}_{+n} + {\cal E} = - \left(A - 1 - n\right)^2
        + {\cal E}, \nonumber \\
  && E^{(A-1,B)}_{-n} + {\cal E} = - \left(B - \case{1}{2} -
        n\right)^2 + {\cal E}, \\
  \mbox{Spectrum of $H_3$:}   && E^{(A-1,B-1)}_{+n} + {\cal E} = - \left(A - 1 -
        n\right)^2 + {\cal E}, \nonumber \\
 && E^{(A-1,B-1)}_{-n} + {\cal E} = - \left(B - \case{3}{2}
        - n\right)^2 + {\cal E}. \label{eq:Scarf-para-comp3} 
\end{eqnarray}
From (\ref{eq:Scarf-para-comp1})--(\ref{eq:Scarf-para-comp3}) we find that when going 
from $H_1$ to $H_2$, one suppresses the ground state of $H_1$ at an energy
$\case{1}{2} \left[\left(B - \case{1}{2}\right)^2 - A^2\right] < 0$. Then when going
from $H_2$ to $H_3$, one suppresses a state of $H_2$ at an energy 
$\case{1}{2} \left[A^2 - \left(B - \case{1}{2}\right)^2\right] > 0$. Such a state is
an excited or the ground state according to whether $A > B + \case{1}{2}$ or $A <
B + \case{1}{2}$. In general, if $A - N < B - \case{1}{2} < A - N + 1$, where $N \in 
\{1, 2, 3, \ldots\}$, then the spectrum of $H_{ps}$ is
\begin{eqnarray}
  E_0 & = & \case{1}{2} \left[\left(B - \case{1}{2}\right)^2 - A^2\right], \qquad d_0 =
       1, \nonumber \\
  E_1 & = & E_0 + 2A - 1, \qquad d_1 = 3, \nonumber \\
  & \vdots & \nonumber \\
  E_{N-1} & = & E_0 + (N - 1)(2A + 1 - N), \qquad d_{N-1} = 3, \nonumber \\
  E_N & = & E_0 + A^2 - \left(B - \case{1}{2}\right)^2, \qquad d_N = 2, \nonumber \\
  & \vdots & \nonumber \\
  E_{N+2p+1} & = & E_0 + (N + p)(2A - N - p), \qquad d_{N+2p+1} = 3, \nonumber
        \\
  && p = 0, 1, \ldots, p_{+max}, \nonumber \\
  E_{N+2p} & = & E_0 + A^2 - \left(B - \case{1}{2} - p\right)^2, \qquad d_{N+2p} = 3,
        \nonumber \\
  && p = 1, 2, \ldots, p_{-max},
\end{eqnarray} 
where $d_i$ ($i = 0$, 1, \ldots, $N-1$, $N$, \ldots, $N+2p+1$, $N+2p$) is the
degeneracy, $p_{+max} = n_{+max} - N$, and $p_{-max} = n_{-max}$.\par
%
%
Keeping in mind the invariance of~(\ref{eq:Scarf-H}) under $A + \case{1}{2}
\leftrightarrow B$, we can also define another set of superpotentials
\begin{eqnarray}
  W'_1(x)  & = & W^{\prime(A,B)}(x), \qquad W'_2(x) = W^{(A,B-1)}(x), \nonumber\\
  c'_1 & = & - c'_2 = \case{1}{2} \left[A^2 - \left(B - \case{1}{2}\right)^2\right].
\end{eqnarray}
Then using~(\ref{eq:Scarf-H-compbis}) and~(\ref{eq:Scarf-H-comp}), we get
\begin{equation}
  H'_1 = H^{(A,B)} + {\cal E}, \qquad H'_2 = H^{(A,B-1)} + {\cal E}, \qquad H'_3 =
         H^{(A-1,B-1)} + {\cal E}, 
\end{equation}
implying the following spectra:
\begin{eqnarray}
  \mbox{Spectrum of $H'_1$:} && E^{(A,B)}_{+n} + {\cal E} = - \left(A - n\right)^2 +
        {\cal E}, \nonumber \\
  && E^{(A,B)}_{-n} + {\cal E} = - \left(B - \case{1}{2} - n
        \right)^2 + {\cal E}, \\
  \mbox{Spectrum of $H'_2$:} && E^{(A,B-1)}_{+n} + {\cal E} = - \left(A - n\right)^2
        + {\cal E}, \nonumber \\
  && E^{(A,B-1)}_{-n} + {\cal E} = - \left(B - \case{3}{2} -
        n\right)^2 + {\cal E}, \\
  \mbox{Spectrum of $H'_3$:}   && E^{(A-1,B-1)}_{+n} + {\cal E} = - \left(A - 1 -
        n\right)^2 + {\cal E}, \nonumber \\
 && E^{(A-1,B-1)}_{-n} + {\cal E} = - \left(B - \case{3}{2}
        - n\right)^2 + {\cal E}. 
\end{eqnarray}
\par
%
%
We thus see that in going from $H'_1$ to $H'_2$, one suppresses an excited state
of $H'_1$ at an energy $\case{1}{2} \left[A^2 - \left(B - \case{1}{2}\right)^2\right]
> 0$. Then when going from $H'_2$ to $H'_3$, one suppresses the ground state of
$H'_2$ at an energy $\case{1}{2} \left[\left(B - \case{1}{2}\right)^2 - A^2\right] < 0$.
If $A - N < B - \case{1}{2} < A - N + 1$, where $N \in \{1, 2, 3, \ldots\}$, then the
spectrum of $H_{ps}$ is the same as in the previous case of~(\ref{eq:Scarf-para-super}),
but the degeneracies are $d_0 = 2$, $d_1 = 3$, \ldots, $d_{N-1} = 3$, $d_N = 1$,
\ldots, $d_{N+2p+1} = 3$, $d_{N+2p} = 3$.\par
%
%
Whenever $A - B + \frac{1}{2}$ goes to an integer, we observe the same collapse of the
double series of energy levels~\cite{bagchi00c} and restoration of the usual PSUSY scheme
as in the two previous subsections.\par
%
%
\section{In pursuit of a complexified SSUSY}

\setcounter{equation}{0}

\subsection{Underlying ideas of SSUSY}

SSUSY is an extended supersymmetric theory having a second-derivative realization of the
differential operators $A$ and $\Ab$~\cite{andrianov93}--\cite{bagchi99}. SSUSY
schemes find interesting applicability to non-trivial quantum mechanical problems, which
include coupled channel problems and those related to transparent matrix potentials.
SSUSY is not guided by a Schr\"odinger form of the Hamiltonian operator, but instead by
a quasi-Hamiltonian $K$, which is a fourth-order differential operator. However, under
certain conditions, $K$ can be related to the square of the Schr\"odinger Hamiltonian:
indeed this feature has been exploited to arrive at models of PSUSY by glueing two
ordinary SUSY systems~\cite{andrianov93}.\par
%
%
Consider supercharges involving second derivatives ($\partial \equiv d/dx$):
\begin{eqnarray}
  \Ap & = & \partial^2 - 2 p(x) \partial + b(x), \label{eq:Ap}\\
  \Am & = & \partial^2 + 2 p(x) \partial + 2 p'(x) + b(x), \label{eq:Am}  
\end{eqnarray}
where $p(x)$ and $b(x)$ are arbitrary functions. Let us introduce the following
operators built out of $\Ap$ and $\Am$:
\begin{equation}
  Q^+ = \left(\begin{array}{cc}
            0 & 0 \\[0.2cm]
            \Am & 0
            \end{array}\right), \qquad
  Q^- = \left(\begin{array}{cc}
            0 & \Ap \\[0.2cm]
            0 & 0
            \end{array}\right). \label{eq:SSUSY-charges}
\end{equation}
In analogy with~(\ref{eq:Hs}), we can think of a quasi-Hamiltonian $K$ defined by
\begin{equation}
  K = Q^+ Q^- + Q^- Q^+. \label{eq:K}
\end{equation}
Clearly $K$ is a fourth-order differential operator.\par
%
%
We can also construct another operator $H$ from two Schr\"odinger-like Hamiltonians
$h^{(1)}$ and $h^{(2)}$:
\begin{eqnarray}
  H & = & \left(\begin{array}{cc}
            h^{(1)} & 0 \\[0.2cm]
            0 & h^{(2)}
            \end{array}\right), \label{eq:SSUSY-H}\\
  h^{(1,2)} & = & - \partial^2 + V^{(1,2)},
\end{eqnarray}
such that $H$ commutes with $Q^{\pm}$:
\begin{equation}
  \left[H, Q^{\pm}\right] = 0. \label{eq:SSUSY-com}
\end{equation}
\par
%
%
{}From (\ref{eq:SSUSY-charges}), (\ref{eq:SSUSY-H}), and (\ref{eq:SSUSY-com}), we are
led to
\begin{equation}
  \Am h^{(1)} = h^{(2)} \Am, \qquad \Ap h^{(2)} = h^{(1)} \Ap. 
  \label{eq:SSUSY-intertwine}
\end{equation}
These are intertwining relationships similar to the supersymmetric
ones~(\ref{eq:SUSY-intertwine}).\par
%
%
Using the representations~(\ref{eq:Ap}) and~(\ref{eq:Am}), we can
exploit~(\ref{eq:SSUSY-intertwine}) to obtain constraints among the functions $p(x)$,
$b(x)$, and the potentials $V^{(1,2)}(x)$:
\begin{eqnarray}
  b & = & - p' + p^2 - \frac{p''}{2p} + \left(\frac{p'}{2p}\right)^2 + \frac{d}{4p^2}, \\
  V^{(1,2)} & = & \mp 2p' + p^2 + \frac{p''}{2p} - \left(\frac{p'}{2p}\right)^2 -
          \frac{d}{4p^2} - a, \label{eq:SSUSY-V-comp} 
\end{eqnarray}
where $d$ and $a$ are integration constants and the primes denote derivatives with
respect to $x$.\par
%
%
We next address to what is known as polynomial SUSY. Here the quasi-Hamiltonian $K$ is
taken to be a quadratic in $H$:
\begin{equation}
  K = H^2 + 2\alpha H + \beta = (H + a)^2 + d,
\end{equation}
where $\alpha$, $\beta$ are constants, and $a = \alpha$, $d = \beta - \alpha^2$. A
PSUSY model can be developed~\cite{andrianov93} by choosing $a = 0$, for which
\begin{equation}
  K = H^2 + d.
\end{equation}
Factorization of $K$ requires $d$ to be a perfect square in the form $d = \frac{c^2}{4}$
for $d > 0$ or $d = - \frac{c^2}{4}$ for $d < 0$. Andrianov {\sl et
al.}~\cite{andrianov95a, andrianov95b} call $d < 0$ a reducible algebra and $d > 0$ an
irreducible one. In the reducible case, we can imagine the existence of an
intermediate Hamiltonian that behaves like a superpartner to both $h^{(1)}$ and
$h^{(2)}$. Alternatively, this triplet of Hamiltonians furnishes a model for PSUSY.\par
%
%
In the following we will be interested in the reducible case only and write
\begin{eqnarray}
  K & = & H^2 - \case{c^2}{4} \nonumber \\
  & = & \left(\begin{array}{cc}
        \left(h^{(1)} + \frac{c}{2}\right) \left(h^{(1)} - \frac{c}{2}\right) & 0 \\[0.2cm]
        0 & \left(h^{(2)} - \frac{c}{2}\right) \left(h^{(2)} + \frac{c}{2}\right) 
        \end{array}\right). \label{eq:K-red1}
\end{eqnarray}
We also know from~(\ref{eq:K}) and~(\ref{eq:SSUSY-charges}) that 
\begin{equation}
  K = \left(\begin{array}{cc}
        \Ap \Am & 0 \\[0.2cm]
        0 & \Am \Ap 
        \end{array}\right). \label{eq:K-red2} 
\end{equation}
Our immediate problem will be to reconcile~(\ref{eq:K-red1}) and~(\ref{eq:K-red2}). To
this end, we factorize $\Ap$ and $\Am$ as
\begin{eqnarray}
  \Ap & = & q^+_1 q^+_2 = (- \partial + W_1) (- \partial + W_2), \nonumber \\
  \Am & = & q^-_2 q^-_1 = (\partial + W_2) (\partial + W_1).
\end{eqnarray}
We thus run into a pair of superpotentials $W_1$ and $W_2$ in SSUSY quite naturally.
Next choosing a constraint
\begin{equation}
  q^+_2 q^-_2 - \case{c}{2} = q^-_1 q^+_1 + \case{c}{2}, \label{eq:SSUSY-constraint}
\end{equation}
we at once see that we can express
\begin{eqnarray}
  \Ap \Am & = & \left(q^+_1 q^-_1 + \case{c}{2} + \case{c}{2}\right) \left(q^+_1
        q^-_1 + \case{c}{2} - \case{c}{2}\right), \nonumber \\
  \Am \Ap & = & \left(q^-_2 q^+_2 - \case{c}{2} - \case{c}{2}\right) \left(q^-_2
        q^+_2 - \case{c}{2} + \case{c}{2}\right). \label{eq:ApAm} 
\end{eqnarray}
Eq.~(\ref{eq:ApAm}) suggests that we can interpret
\begin{equation}
  h^{(1)} = q^+_1 q^-_1 + \case{c}{2}, \qquad h^{(2)} = q^-_2 q^+_2 - \case{c}{2}. 
  \label{eq:SSUSY-H-comp}
\end{equation}
Hence (\ref{eq:K-red1}) and (\ref{eq:K-red2}) can be reconciled.\par
%
%
{}From (\ref{eq:SSUSY-H-comp}), we further have
\begin{eqnarray}
  h^{(1)} & = & (- \partial + W_1) (\partial + W_1) + \frac{c}{2} \nonumber \\
  & = & - \partial^2 + V^{(1)}(x), \\ 
  h^{(2)} & = & (\partial + W_2) (- \partial + W_2) - \frac{c}{2} \nonumber \\
  & = & - \partial^2 + V^{(2)}(x), 
\end{eqnarray}
reflecting
\begin{equation}
  V^{(1)}(x) = W_1^2 - \frac{dW_1}{dx} + \frac{c}{2}, \qquad V^{(2)}(x) = W_2^2 +
  \frac{dW_2}{dx} - \frac{c}{2}. \label{eq:SSUSY-V-compbis}  
\end{equation}
\par
%
%
To tie up, we confront~(\ref{eq:SSUSY-V-comp}) with the
expressions~(\ref{eq:SSUSY-V-compbis}). The results for $a = 0$ are 
\begin{equation}
  W_1 = - \frac{2p' + c}{4p} + p, \qquad W_2 = \frac{2p' + c}{4p} + p. 
  \label{eq:SSUSY-super}
\end{equation}
We are thus led to explicit forms of the two superpotentials $W_1$ and $W_2$ in terms
of the function $p(x)$ only.\par
%
%
Note that there exists, notionally, an intermediate Hamiltonian $h$, which is superpartner
to both $h^{(1)}$ and $h^{(2)}$:
\begin{equation}
  h^{(1)} = q^+_1 q^-_1 + \case{c}{2}, \qquad h = q^-_1 q^+_1 + \case{c}{2}, \qquad
  h^{(2)} = q^-_2 q^+_2 - \case{c}{2}.
\end{equation}
Due to the constraint~(\ref{eq:SSUSY-constraint}), we can express $h$ as
\begin{equation}
  h = q^-_1 q^+_1 + \case{c}{2} = q^+_2 q^-_2 - \case{c}{2}. 
\end{equation}
The constraint~(\ref{eq:SSUSY-constraint}), when exposed in terms of the
superpotentials $W_1$ and $W_2$, reads
\begin{equation}
  W_2^2 - W_1^2 - \frac{dW_1}{dx} - \frac{dW_2}{dx} = c. \label{eq:constraint}  
\end{equation}
Eq.~(\ref{eq:constraint}) coincides with~(\ref{eq:PSUSY-constraint}).\par
%
%
\subsection{PT-symmetric oscillator potential}

{}First of all, we notice that $h^{(1)}$, $h$, $h^{(2)}$ defined above go over to $H_1$,
$H_2$, $H_3$ of~(\ref{eq:Hps}), respectively, provided we identify $q^+_1$, $q^+_2$
with $\Ab_1$, $\Ab_2$, and the constants $c_1$, $c_2$ with $c/2$, $-c/2$,
respectively. The latter certainly hold since in~(\ref{eq:PSUSY-Hcomp}) we have taken
$c_1 + c_2 = 0$.\par
%
%
Setting now $c = - 4\alpha$ and $p(x) = x - {\rm i} \delta$, it is trivial to see that $W_1$
and $W_2$ in~(\ref{eq:SSUSY-super}) get complexified and are mapped to the expressions
of $W^{(\alpha)}(x)$ and $W^{\prime(\alpha-1)}(x)$ given by~(\ref{eq:osc-super1})
and~(\ref{eq:osc-super2}), respectively.\par
%
%
On the other hand, if we set $c = + 4\alpha$, then $W_1$ and $W_2$ 
in~(\ref{eq:SSUSY-super}) are mapped to $W^{\prime(\alpha)}(x)$ and 
$W^{(\alpha+1)}(x)$ given by~(\ref{eq:osc-super2}) and~(\ref{eq:osc-super1}),
respectively.\par
%
%
Concerning the constraint relation~(\ref{eq:constraint}), we observe that it holds both
for $c = - 4\alpha$ and $c = + 4\alpha$ if the corresponding expressions for $W_1$ and
$W_2$ are plugged in.\par
%
%
{}Finally, from the point of view of SSUSY we can associate with PSUSY
Hamiltonian~(\ref{eq:Hps}) two distinct SUSY Hamiltonians given by
\begin{equation}
  H^{(1)}_s = \left(\begin{array}{cc}
              \Ab_1 A_1 + \frac{c}{2} & 0 \\
              0 & A_1 \Ab_1 + \frac{c}{2}
              \end{array}\right), \qquad
  H^{(2)}_s = \left(\begin{array}{cc}
              \Ab_2 A_2 - \frac{c}{2} & 0 \\
              0 & A_2 \Ab_2 - \frac{c}{2}
              \end{array}\right), \label{eq:osc-SSUSY-H} 
\end{equation}
where $c = \pm 4\alpha$. Conversely, we could arrive at the $p=2$ PSUSY form for the
Hamiltonian by glueing $H^{(1)}_s$ and $H^{(2)}_s$ given
by~(\ref{eq:osc-SSUSY-H}).\par
%
%
In the following we show that the results of the PT-symmetric generalized P\"oschl-Teller
and Scarf II potentials are similar to those just obtained for the PT-symmetric oscillator
one.\par
%
%
\subsection{PT-symmetric generalized P\"oschl-Teller potential}

With the first choice of superpotentials coming from the analysis carried out for the
PT-symmetric generalized P\"oschl-Teller problem earlier, namely
\begin{eqnarray}
  W_1 & = & W^{(A,B)} = \left(B - \case{1}{2}\right) \coth \tau - \left(A +
        \case{1}{2}\right) \cosech \tau, \nonumber \\
  W_2 & = & W^{\prime(A,B-1)} = A \coth \tau - (B - 1) \cosech \tau, 
        \label{eq:PT-SSUSY-super}
\end{eqnarray}
it is easy to see that (\ref{eq:PT-SSUSY-super}) fits into the
scheme~(\ref{eq:SSUSY-super}) for the combination
\begin{eqnarray}
  p(x) & = & \case{1}{2} \left(A + B - \case{1}{2}\right) (\coth \tau - \cosech \tau),
           \nonumber \\
  c  & = & \left(A + B - \case{1}{2}\right) \left(A - B + \case{1}{2}\right). 
           \label{eq:PT-p}
\end{eqnarray}
\par
%
%
If we consider instead the second choice
\begin{eqnarray}
  W_1 & = & W^{\prime(A,B)} = A \coth \tau - B \cosech \tau, \nonumber \\
  W_2 & = & W^{(A-1,B)} = \left(B - \case{1}{2}\right) \coth \tau - \left(A -
        \case{1}{2}\right) \cosech \tau, 
\end{eqnarray}
we need only to interchange $A$ and $B - \case{1}{2}$ in~(\ref{eq:PT-p}). Thus $p(x)$
is left unchanged while $c$ just changes sign:
\begin{eqnarray}
  p(x) & = & \case{1}{2} \left(A + B - \case{1}{2}\right) (\coth \tau - \cosech \tau),
           \nonumber \\
  c  & = & - \left(A + B - \case{1}{2}\right) \left(A - B + \case{1}{2}\right). 
\end{eqnarray}
\par
%
%
\subsection{PT-symmetric Scarf II potential}

Here our first choice of superpotentials comes from~(\ref{eq:Scarf-super1})
and~(\ref{eq:Scarf-super2}):
\begin{eqnarray}
  W_1 & = & W^{(A,B)} = A \tanh x + {\rm i} B \sech x, \nonumber \\
  W_2 & = & W^{\prime(A-1,B)} = \left(B - \case{1}{2}\right) \tanh x + {\rm i} \left(
        A - \case{1}{2}\right) \sech x. \label{eq:Scarf-SSUSY-super}
\end{eqnarray}
On seeking consistency with~(\ref{eq:SSUSY-super}), we are led to the solutions
\begin{eqnarray}
  p(x) & = & \case{1}{2} \left(A + B - \case{1}{2}\right) (\tanh x + {\rm i} \sech x),
           \nonumber \\
  c  & = & - \left(A + B - \case{1}{2}\right) \left(A - B + \case{1}{2}\right). 
\end{eqnarray}
\par
%
%
The second choice of superpotentials pertains to
\begin{eqnarray}
  W_1 & = & W^{\prime(A,B)} = \left(B - \case{1}{2}\right) \tanh x + {\rm i} \left(
        A + \case{1}{2}\right) \sech x, \nonumber \\
  W_2 & = & W^{(A,B-1)} = A \tanh x + {\rm i} (B - 1) \sech x.
\end{eqnarray}
This corresponds to an interchange of $A$ and $B - \case{1}{2}$ in the first
choice~(\ref{eq:Scarf-SSUSY-super}). The function $p(x)$ remains the same but $c$
changes sign:
\begin{eqnarray}
  p(x) & = & \case{1}{2} \left(A + B - \case{1}{2}\right) (\tanh x + {\rm i} \sech x),
           \nonumber \\
  c  & = & \left(A + B - \case{1}{2}\right) \left(A - B + \case{1}{2}\right). 
\end{eqnarray}
\par
%
%
\section{Summary}

To summarize our results, we note that in all the three potentials considered by us,
namely the PT-symmetric harmonic oscillator, generalized P\"oschl-Teller, and Scarf II
potentials, we found order-two PSUSY and SSUSY appropriate mediums to account for
their double series of energy levels. Taking cue from the SUSY results, we found possible to
confront the expressions for the relevant superpotential by an appropriate series of
energy levels. These superpotentials, in turn, not only allow developing PSUSY models, but
also adjust nicely with the constraint relations relevant to the SSUSY construction. In this
way, the potentials considered by us can be interpreted in terms of PSUSY and SSUSY
schemes.\par
%
%
\section*{Acknowledgments}

One of us (S.\ M.) thanks the Council of Scientific and Industrial Research, New Delhi for
financial support.\par
%
%

\end{document}